# Nonlinear Nanophotonic Circuitry:
# Tristable and Astable Multivibrators and Chaos Generator

*Pujuan Ma[1,2], Lei Gao[1,#], Pavel Ginzburg[3,4], and Roman E. Noskov[3,4,§]*

[1]School of Physical Science and Technology & Jiangsu Key Laboratory of Thin Film of Soochow University, Collaborative Innovation Center of Suzhou Nano Science and Technology, Soochow University, Suzhou 215006, China.

[2] Shandong Provincial Engineering and Technical Center of Light Manipulations & Shandong Provincial Key Laboratory of Optics and Photonic Device, School of Physics and Electronics, Shandong Normal University, Jinan 250014, China

[3] Department of Electrical Engineering, Tel Aviv University, Ramat Aviv, Tel Aviv 69978, Israel

[4] Light-Matter Interaction Centre, Tel Aviv University, Ramat Aviv, Tel Aviv, 69978, Israel

[#]e-mail: leigao@suda.edu.cn

[§]e-mail: nanometa@gmail.com

**Abstract:** The concept of lumped optical nanoelements (or metactronics), wherein nanometer-scale structures act as nanoinductors, nanocapacitors and nanoresistors, has attracted a great deal of attention as a simple toolbox for engineering different nanophotonic devices in analogy with microelectronics. While recent studies of the topic have been predominantly focused on linear functionalities, nonlinear dynamics in microelectronic devices plays a crucial role and provides a majority of functions, employed in modern applications. Here, we extend the metactronics paradigm and add nonlinear dynamical modalities to those nanophotonic devices that have never been associated with optical nanoantennas. Specifically, we show that nonlinear dimer nanoantennae can operate in the regimes of tristable and astable multivibrators as well as chaos generators. The physical mechanism behind these modalities relies on the Kerr-type nonlinearity of nanoparticles in the dimer enhanced by a dipolar localized surface plasmon resonance. This allows one to provide a positive nonlinear feedback at moderate optical intensities, leading to the desired dynamical behavior via tuning the driving field parameters. Our findings shed light on a novel class of nonlinear nanophotonic devices with a tunable nonlinear dynamical response.

## 1. Introduction

In microelectronics, nonlinear circuits find applications in a variety of devices such as timers, logic gates, frequency dividers, data storage, signal generators and many others [1]. Being one of the cornerstones for nonlinear electronic circuits, multivibrators represent a broad class of two-steady-state systems with different kinds of switching behavior [2].

In monostable multivibrators one of the steady states is stable while another is quasi-stable (meta-stable). A signal pulse triggers a circuit into the quasi-stable state, which is kept for a set time period, after which the system spontaneously (without any signal pulse) returns to its initial stable state. Monostable multivibrators are generally used to convert short sharp pulses into wider ones with a fixed duration for timing applications.

Bistable multivibrators (also known as flip-flops) are characterized by two stable steady states. This circuit changes its state in response to a triggering pulse only. Due to the possibility to store a bit of information, such systems are widely exploited in computer memory and other digital electronic circuits nowadays.

In an astable multivibrator (also known as a relaxation oscillator) neither steady state is stable. Such system operates in the mode of periodic self-oscillations, giving rise to a non-sinusoidal repetitive output signal. Relaxation oscillators are employed to create low frequency signals for function generators, electronic beepers, inverters, switching power supplies etc.

Finally, one of the most interesting concepts, realized in nonlinear electronic circuitry, is generation of deterministic chaos when the system demonstrates unpredictable dynamical behavior [3]. The classic example of such scenario is Chua's circuit [4]. In optics chaotic behavior is well-studied for multimode lasers [5]. Since chaotic regimes exhibit very complicated dynamic features, they have found applications in secure data processing for confusion and diffusion operations as well as true random numbers generation for security keys [6–8].

These devices have been extensively developed and widely used in kHz, MHz and GHz frequency domains as a result of comparatively easy design and analysis based on the concept of linear (resistors, inductors, capacitors) and nonlinear (e.g., diodes, triodes, operational amplifiers) lumped elements. However, modern telecommunication and information technologies require much higher operational frequencies where lumped elements cannot be obtained by straightforward scaling down the size, since material dispersion significantly alters properties of metals and dielectrics and the critical latency in interconnections appears [9–11]. Hence, one can pose the reasonable question: How to engineer and analyze dynamical nonlinear circuits in infrared (IR) and visible ranges? Here, we introduce a general approach based on using intrinsic Kerr-type nonlinearity of optical lamped nanoelements to resolve this issue. Specifically, we consider a dimer nanoantenna that is made from a pair of identical nanoparticles possessing a strong cubic susceptibility enhanced by a resonant response. In practice, such requirement is met for a variety of plasmonic materials such as silver, gold and graphene [12–15] either in optical or infrared spectral range. Relying on equivalent capacitance, inductance and resistance of nanoparticles, we derive a model describing the nonlinear dynamical behavior of the dimer and reveal the regimes of tristable multivibrator, associated with switching between two stable steady states via a quasi-stable equilibrium, astable multivibrator and chaos generator. All these functionalities co-exist in a single nanoantenna and triggering between them can be realized via adjusting the driving field intensity, frequency, and orientation. Additionally, we show that, despite the homogeneous field excitation, every kind of dynamical behavior can be accompanied by the system symmetry breaking, associated with an asymmetric response of nanoparticle dipole moments.

## 2. Theoretical Framework

To rigorously analyze the system dynamics, we describe the subwavelength nanoparticles in the dimer within the point-dipole approximation and assume the dimer excitation by a plane wave with the frequency close to the nanoparticle's dipolar resonance ($\omega_0$) (Fig. 1(a)). For the sake of clarity, here we consider graphene-wrapped dielectric nanoparticles of a spherical shape while the general model derivation for any nanoparticle shape and composition can be found in Supplementary Materials. Following Refs. [9,16], we treat each nanoparticle as a pair of equivalent parallel RLC circuits obeying the Kirchhoff's current law in the frequency domain as follows (Fig. 1(b))

$$\begin{pmatrix} Z^{-1}(\omega) & 0 & -i\omega\Delta C_\parallel & 0 \\ 0 & Z^{-1}(\omega) & 0 & -i\omega\Delta C_\perp \\ -i\omega\Delta C_\parallel & 0 & Z^{-1}(\omega) & 0 \\ 0 & -i\omega\Delta C_\perp & 0 & Z^{-1}(\omega) \end{pmatrix} \begin{pmatrix} U_1^\parallel \\ U_1^\perp \\ U_2^\parallel \\ U_2^\perp \end{pmatrix} = \begin{pmatrix} I_{ext}^\parallel \\ I_{ext}^\perp \\ I_{ext}^\parallel \\ I_{ext}^\perp \end{pmatrix}, \quad (1)$$

where $Z^{-1}(\omega) = \frac{1}{R(\omega)} - \frac{1}{i\omega L(\omega)} - i\omega C$ is the nanoparticle impedance ($\exp(-i\omega t)$ time dependence is assumed). The resistance, inductance, and capacitance for a graphene-wrapped nanoparticle with the permittivity of the core $\varepsilon_{core}$, the graphene conductivity $\sigma_{NL}$ and the radius $a$ can be written as [9]

$$R(\omega) = \frac{2\operatorname{Re}\sigma_{NL}(\omega)}{\pi\left(\omega a\varepsilon_0\varepsilon_{core} + 2\operatorname{Im}\sigma_{NL}(\omega)\right)^2}, \quad L(\omega) = -\frac{1}{\omega^2\left(\omega a\varepsilon_0\varepsilon_{core} + 2\operatorname{Im}\sigma_{NL}(\omega)\right)}, \quad C = 2\pi a\varepsilon_0\varepsilon_h,$$

where $\varepsilon_h$ and $\varepsilon_0$ are the permittivities of the host matrix and the vacuum, respectively. We assume the Kerr-like nonlinearity of the graphene layer that can be described as $\sigma_{NL}(\omega) = \sigma_L(\omega) + \Delta\sigma$, where the linear term $\sigma_L(\omega)$ fits the Drude model and the nonlinear correction $\Delta\sigma = \sigma^{(3)}|\mathbf{E}_{in}|^2$ includes the local electric field inside the nonlinear media $\mathbf{E}_{in}$ and the cubic conductivity $\sigma^{(3)}$ (see Supplementary Materials for details). $U_{1,2}^{\parallel,\perp} = \frac{p_{1,2}^{\parallel,\perp}}{4\pi\varepsilon_0\varepsilon_h a^2}$ is the effective potential difference between the upper and the lower hemispherical surfaces of nanoparticles [9], and $p_{1,2}$ is the nanoparticle dipole moment. $I_{ext}^{\parallel,\perp} = -i\omega\varepsilon_0(\varepsilon_{core} + 2i\sigma_{NL}[\omega a\varepsilon_0]^{-1} - \varepsilon_h)\pi a^2 E_{ext}^{\parallel,\perp}$ is the effective displacement current driving the system [9], and $E_{ext}$ is the electric field of a plane wave. The indexes '$\parallel$' and '$\perp$' stand for the longitudinal and the transversal components with respect to the system symmetry axis. The dipole-dipole interaction between nanoparticles depends on the orientation of the external field as well as the distance between their centers $d$ and can be described via additional capacitances $\Delta C_{\parallel} = 2\pi\varepsilon_0\varepsilon_h a(a/d)^3$ and $\Delta C_{\perp} = -4\pi\varepsilon_0\varepsilon_h a(a/d)^3$. It is important to note that the point-dipole approximation is valid as long as $d \geq 3a$ [17]. These terms result in an anisotropic response of the dimer so that it can be considered as 4 equivalent RLC circuits coupled linearly via capacitances, as shown in Fig. 1(b).

In our model we neglect quantum finite-size and nonlocality effects which appear when the size of nanoparticles gets smaller than 10 nm [18,19].

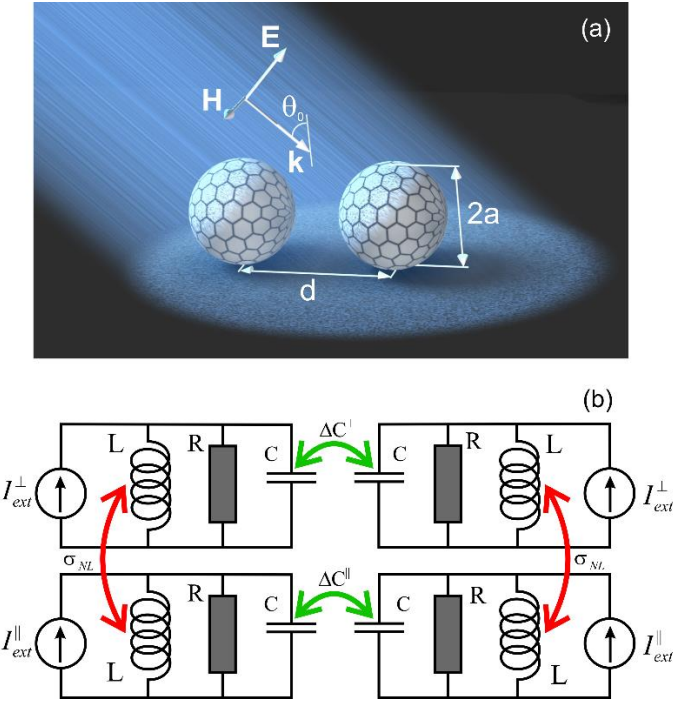

**Fig. 1.** (a) Schematics of a graphene-wrapped nanodimer illuminated by a plane wave. (b) The equivalent scheme of a nonlinear nanodimer presented in terms of lumped elements. Close to the resonance each nanoparticle acts as a pair of RLC-oscillators corresponding to transversal and longitudinal surface plasmonic oscillations with respect to the nanodimer axis (denoted by the superscripts). The interparticle coupling is predominantly capacitive, and the nonlinearity resides in the nanoparticle inductances yielding transversal and longitudinal nonlinear coupling inside each nanoparticle. The driving electric field projections are associated with the effective displacement currents $I_{ext}^{\parallel,\perp}$.

Within the framework of the slowly varying amplitude approximation, we employ the dispersion relation method [20,21] that assumes that nonlinearity, losses, frequency detuning from resonance, and broadening of the nanoparticle polarization spectrum are accounted for in the first order of the perturbation theory, i.e., $\Delta\sigma \ll \sigma_L$, $\mathrm{Im}\,\sigma_L \gg \mathrm{Re}\,\sigma_L$, $(\omega-\omega_0)/\omega_0 \ll 1$. The physical interpretation is as follows: each nanoparticle acts as a resonantly excited oscillator with slow (in comparison with the oscillation period of the driving field) inertial response that allows us to treat their dynamical response in terms of quantities averaged over the oscillation period. Importantly, these assumptions imply that the nonlinearity resides only in the nanoparticle inductances, giving rise to inductive coupling for both components of the nanoparticle polarization (Fig. 1(b)). We present $Z^{-1}(\omega)$ and $-i\omega\Delta C_{\parallel,\perp}$ as a Taylor series in the vicinity of $\omega_0$

$$Z^{-1}(\omega) \approx \partial_{\Delta\sigma}Z^{-1}\big|_{\substack{\text{Im}\varepsilon_L=0 \\ \omega=\omega_0}} \Delta\sigma + iZ^{-1}\big|_{\substack{\Delta\sigma=0 \\ \omega=\omega_0}} + \partial_\omega Z^{-1}\big|_{\substack{\Delta\sigma=0 \\ \text{Im}\varepsilon_L=0 \\ \omega=\omega_0}} \left(\Delta\omega + i\frac{d}{dt}\right),$$

$$-i\omega\Delta C_{\parallel,\perp} \approx -i\omega_0 \Delta C_{\parallel,\perp}.$$

and substitute these expressions into Eq. (1), which leads to a set of coupled equations for the slowly varying amplitudes of the nanoparticle dipole moments $\mathbf{P}_1$ and $\mathbf{P}_2$, written in dimensionless units as follows:

$$\begin{cases} i\dfrac{dP_1^P}{d\tau} + \left[\left|P_1^P\right|^2 + \left|P_1^\perp\right|^2 + i\gamma + \Omega\right]P_1^P + G^P P_2^P = E^\parallel \\ i\dfrac{dP_1^\perp}{d\tau} + \left[\left|P_1^P\right|^2 + \left|P_1^\perp\right|^2 + i\gamma + \Omega\right]P_1^\perp + G^\perp P_2^\perp = E^\perp \\ i\dfrac{dP_2^P}{d\tau} + \left[\left|P_2^P\right|^2 + \left|P_2^\perp\right|^2 + i\gamma + \Omega\right]P_2^P + G^P P_1^P = E^\parallel \\ i\dfrac{dP_2^\perp}{d\tau} + \left[\left|P_2^P\right|^2 + \left|P_2^\perp\right|^2 + i\gamma + \Omega\right]P_2^\perp + G^\perp P_1^\perp = E^\perp \end{cases} \qquad (2)$$

In this formulation, the normalization is given by

$$P_{1,2}^{\parallel,\perp} = 4\pi\varepsilon_0\varepsilon_h a^2 N_p U_{1,2}^{\parallel,\perp},$$

$$E^{\parallel,\perp} = \left[-i\omega_0\varepsilon_0(\varepsilon_{core} + 2i\sigma_L(\omega_0)[\omega_0 a\varepsilon_0]^{-1} - \varepsilon_h)\pi a^2\right]^{-1} N_E I_{ext}^{\parallel,\perp},$$

$$N_p = \left[\partial_{\Delta\varepsilon}Z^{-1}(\omega_0)\{\partial_\omega Z^{-1}(\omega_0)\omega_0\}^{-1}\psi\alpha^{-2}(\omega_0)\sigma^{(3)}\right]^{1/2},$$

$$N_E = \left[\partial_{\Delta\varepsilon}Z^{-1}(\omega_0)\{\partial_\omega Z^{-1}(\omega_0)\omega_0\}^{-3}\psi\alpha^{-2}(\omega_0)\sigma^{(3)}\right]^{1/2},$$

$$\gamma = R^{-1}(\omega_0)\cdot\left[\partial_\omega Z^{-1}(\omega_0)\cdot\omega_0\right]^{-1},$$

$$G^{\parallel,\perp} = -i\Delta C_{\parallel,\perp}^{-1}\cdot\left[\partial_\omega Z^{-1}(\omega_0)\right]^{-1}.$$

Here $G^{P,\perp}$ describes the dipole-dipole coupling between nanoparticles, $\Omega = (\omega-\omega_0)/\omega_0$ is the driving frequency detuning from the resonant value, $\gamma$ accounts for the thermal and radiation losses of the nanoparticles, $\tau = \omega_0 t$ is the dimensionless time, $\psi$ is the local field enhancement factor at resonance, $\alpha$ is the nanoparticle polarizability (see Supplementary Materials), and $E^{\parallel,\perp} = E_p^{\parallel,\perp} + E_s^{\parallel,\perp}$ are the components of the slow varying amplitude of the driving optical field, which includes pumping continuous-wave (cw) background radiation $E_p^{\parallel,\perp}$ and signal pulse $E_s^{\parallel,\perp}$ contributions. Since the interparticle distance is much smaller than the light wavelength, we neglect the phase delay in the local field acting on different nanoparticles. We assume that the background field slowly grows and saturates at the level $E_0$ as $E_p = (2E_0 / \pi)\arctan(\tau / \Delta \tau)$ ($\Delta \tau$ is the characteristic saturation time), and the phase-locked signal pulse follows the Gaussian shape $E_s = E_{peak} \exp(-(\tau - \tau_0)^2 / \delta^2)$, where $E_{peak}$ is the peak amplitude of the pulse, $\tau_0$ defines the pulse temporal localization, and $\delta$ is the pulse half-width. Such form of stimuli is convenient to drive different dynamical regimes, and, in practice, it is common for coherent control over excitations in nanostructures [22–24].

To illustrate the dynamical behavior of the nanodimer in the far field, we also introduce the total scattering cross-section as follows [25]

$$\sigma_{sc} = \frac{\omega^4 N_p^2}{16\pi^2 \varepsilon_0^2 \varepsilon_h^2 c^4 E^2} \int_0^{2\pi} \int_0^{\pi} \left( \left[ |P_1^\perp|^2 + |P_2^\perp|^2 + 2|P_1^\perp|^2 |P_2^\perp|^2 \cos(\Delta\Psi_\perp + kd\sin\vartheta\sin\varphi) \right] \sin^2 \vartheta + \left[ |P_1^\parallel|^2 + |P_2^\parallel|^2 + 2|P_1^\parallel|^2 |P_2^\parallel|^2 \cos(\Delta\Psi_\parallel - kd\sin\vartheta\sin\varphi) \right] \left[ 1 - \sin^2 \vartheta \sin^2 \varphi \right] \right) \sin\vartheta \, d\varphi \, d\vartheta,$$

where $\varphi$ and $\theta$ are the spherical azimuthal and polar angles, respectively, $\Delta\Psi_{\parallel,\perp}$ denotes an internal phase shift between $P_1^{\parallel,\perp}$ and $P_2^{\parallel,\perp}$, and $k$ is the wave number.

Eq. (2) is of the Duffing type that possesses a very rich dynamical behavior [26]. Hereinafter, we systematically analyze it and reveal novel dynamical functionalities that have never been associated with optical nanoantennas.

## 3. Results and Discussion

Our model represented by Eq. (2) is universal, capturing dynamical features for a broad class of nonlinear nanophotonic systems. Without loss of generality we provide quantitative estimations of parameters for the following configuration: a pair of identical spherical graphene-wrapped nanoparticles made of $BaF_2$. We have chosen such configuration since graphene is a promising highly nonlinear plasmonic material [27–29] and $BaF_2$ is transparent and almost dispersion-free in the middle infrared domain where graphene demonstrates plasmonic resonances (see Supplementary Materials for details). In experiment, graphene-wrapped nanospheres can be obtained by using layer-by-layer self-assembly or precursor-assisted chemical vapor deposition [30–32]. The nanoparticle radius and the center-to-center distance are $a = 100$ nm and $d = 300$ nm. We adjusted the dimer parameters to obtain resonance at $\hbar\omega_0 = 0.133$ eV, which corresponds to the wavelength of a $CO_2$ laser at 9.32 μm.

The general steady-state solution of Eq. (2) can be presented as,

$$\begin{cases} \left[\left|P_1^P\right|^2 + \left|P_1^\perp\right|^2 + i\gamma + \Omega\right]P_1^P + G^P P_2^P = E_0 \cos(\theta_0) \\ \left[\left|P_1^P\right|^2 + \left|P_1^\perp\right|^2 + i\gamma + \Omega\right]P_1^\perp + G^\perp P_2^\perp = E_0 \sin(\theta_0) \\ \left[\left|P_2^P\right|^2 + \left|P_2^\perp\right|^2 + i\gamma + \Omega\right]P_2^P + G^P P_1^P = E_0 \cos(\theta_0) \\ \left[\left|P_2^P\right|^2 + \left|P_2^\perp\right|^2 + i\gamma + \Omega\right]P_2^\perp + G^\perp P_1^\perp = E_0 \sin(\theta_0) \end{cases}, \quad (3)$$

where $\theta_0$ is the angle of incidence (Fig. 1(a)). This relation includes two kinds of the steady states: symmetric ones with $P_1^P = P_2^P = P_0^P$ and $P_1^\perp = P_2^\perp = P_0^\perp$ and asymmetric ones with $P_1^P \neq P_2^P$ and $P_1^\perp \neq P_2^\perp$. The symmetric solution can be obtained from

$$\begin{cases} \left[\left|P_0^P\right|^2 + \left|P_0^\perp\right|^2 + i\gamma + \Omega\right]P_0^P + G^P P_0^P = E_0 \cos(\theta_0) \\ \left[\left|P_0^P\right|^2 + \left|P_0^\perp\right|^2 + i\gamma + \Omega\right]P_0^\perp + G^\perp P_0^\perp = E_0 \sin(\theta_0) \end{cases}. \quad (4)$$

In the scalar case, i.e., for $\vartheta_0 = 0$ and $\vartheta_0 = \pi/2$, this set has a bistable solution for $\Omega < -\text{Re}\, G^{P,\perp} - \sqrt{3}\left|\gamma - \text{Im}\, G^{P,\perp}\right|$. Due to different dipole-dipole coupling for the dipoles oriented in the longitudinal and transverse directions (expressed via $G^P$ and $G^\perp$), the regions of bistability for $P_0^P$ and $P_0^\perp$ are shifted, leading to tri-stability in the vector case when $\vartheta_0 \neq \{0, \pi/2\}$ (compare the green curves in Figs. 2 (a), (b), and (c)).

Importantly, appearance of asymmetric steady states in presence of symmetric stimuli stems from the nonlinear feedback. It does not exist in the linear case, as follows from Eq. (3). Using a Newton iteration scheme, we find the branches corresponding to this solution and plot them together with the symmetric ones in Figs. 2 (a), (b), and (c).

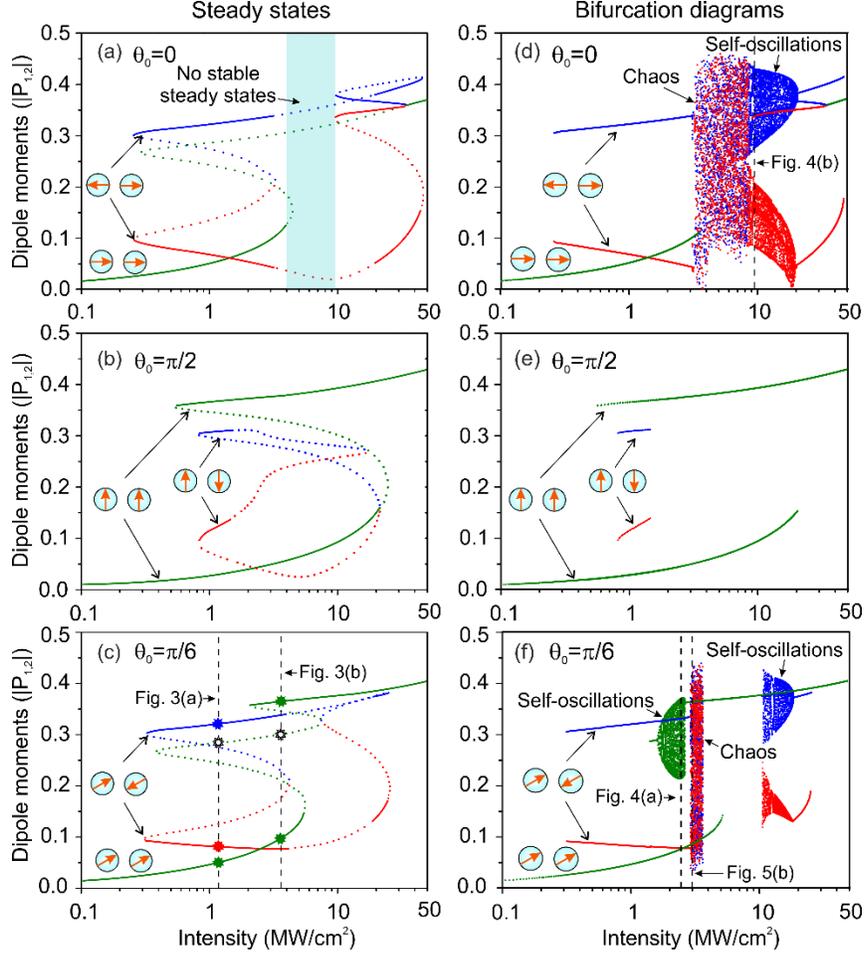

**Fig. 2.** Characterization of the steady states and nonlinear dynamics in the nanodimer consisting of graphene-wrapped $BaF_2$ nanoparticles. Panels (a), (b) and (c) show the steady states for the absolute values of the normalized nanoparticle dipole moments vs light intensity at different incident angles of the plane wave: (a) $\theta_0 = 0$, (b) $\theta_0 = \pi/2$, and (c) $\theta_0 = \pi/6$. Continuous and dotted curves mark stable and unstable branches identified by the linear stability analysis. Symmetric states with equal nanoparticle dipole moments are denoted by green lines. Asymmetric states are shown by blue and red lines, corresponding to the dipole moments of the first and the second nanoparticle. Panels (d), (e) and (f) show bifurcation diagrams obtained for the same parameters as in (a), (b) and (c). In addition to the stable steady states, there appear the regions with regular (self-oscillations) and stochastic dynamics. All plots are obtained for $\Omega = -0.1$. The dashed lines denote the saturation intensities for temporal dynamical responses shown in Figs. 3, 4, and 5. In (c) the lower green stars mark the initial system steady states, the empty white stars show the quasi-stable equilibrium points, and the blue and red stars as well as the upper green star indicate the eventual steady states from dynamical behaviors in Fig. 3.

Our system has eight degrees of freedom corresponding to $P_{1,2}^{P,\perp}$ and their complex conjugates. In order to examine stability of equilibria in this 8-dimensional space, we perform the linear stability analysis and find eight eigen-values of the Jacobian matrix of Eq. (3) (see Supplemental Materials for details). The stability of equilibria is determined by the sign of the real parts for these eigen-values: once the real part for, at least, one eigen-value is positive, the corresponding steady state is unstable. Such unstable equilibria are marked by dotted curves in Figs. 2 (a), (b), and (c). Remarkably, one can distinguish regions with one, two and three stable steady states belonging to symmetric and asymmetric solutions in different combinations. Furthermore, in Fig. 2 (a) there appears quite an untrivial situation: for the domain of light intensities from 3.1 to 8.8 MW cm$^{-2}$ no stable state (neither symmetric nor asymmetric) exists.

To unravel the dynamical behavior of our nanoantenna, we make a series of separate simulations of Eq. (2) varying the cw background saturated optical intensity and the peak intensity of the signal pulse. Having omitted transitional processes (i.e., when the background field has saturated and the pulse has elapsed), we record the absolute values of the nanoparticle dipole moments after every fixed time interval and plot them for every saturated light intensity in Figs. 2 (d), (e), and (f) (see also Supplemental Materials for an angle dependent analysis). The recording time $\tau_{rec} = 4 \times 10^4$ and the duration of the time interval $\tau_{int} = 72$ have been adjusted to cover features of the dynamical processes. In these bifurcation diagrams, the stable steady states correspond to continuous branches, and they are in absolute agreement with the results of the linear stability analysis, shown in Figs. 2 (a), (b), and (c). Moreover, we discover the regions with self-oscillations and chaotic dynamics. Specifically, in the case when no stable state exists (Fig. 2(a)) we get a chaotic regime accompanied by symmetry breaking, i.e. an asymmetric response of nanoparticle dipole moments (Fig. 2 (d)). It is interesting to consider a transition to this regime. If one starts decreasing the light intensity for the stable asymmetric branch from 46 MW cm$^{-2}$, firstly one can observe the Hopf bifurcation at 19 MW cm$^{-2}$ associated with appearance of a stable limiting circle from a stable steady state via losing its stability. Then at 8.8 MW cm$^{-2}$ we observe in numerical simulations that the period doubling bifurcation results in the chaotic attractor (not shown). Notably, a chaotic attractor also can exist separately from the self-oscillations as shown in Fig. 2 (f).

Additionally, the linear stability analysis allowed us to identify the steady state branch characterized by 1 unstable and 7 stable dimensions (marked by the empty stars in Fig. 2(c)). Such equilibria can be treated as *quasi-stable* since repulsion from this equilibrium point becomes significant only when the system state approaches its vicinity along all other stable dimensions. The set time the system keeps this equilibrium is related with the values of the real parts of the eigen-values of the Jacobian matrix for the stable dimensions that can be calculated numerically. Roughly the set time can be evaluated via a sum of the real parts of the eigen-values of the Jacobian matrix for all stable dimensions. Therefore, the number of stable dimensions is important.

After this set time the system leaves the quasi-stable state spontaneously without any signal pulse as a result of repulsion action along an unstable dimension. Its location between stable branches (Fig. 2(c)) opens an interesting possibility to trigger the nanoantenna between them via the quasi-stable state. Since this operation, in fact, represents a combination of responses for bi- and monostable multivibrators, we refer to it as a tristable multivibrator. Next, we consider two cases meeting this scenario for which the saturated optical intensities are denoted in Fig. 2(c) by the dashed lines.

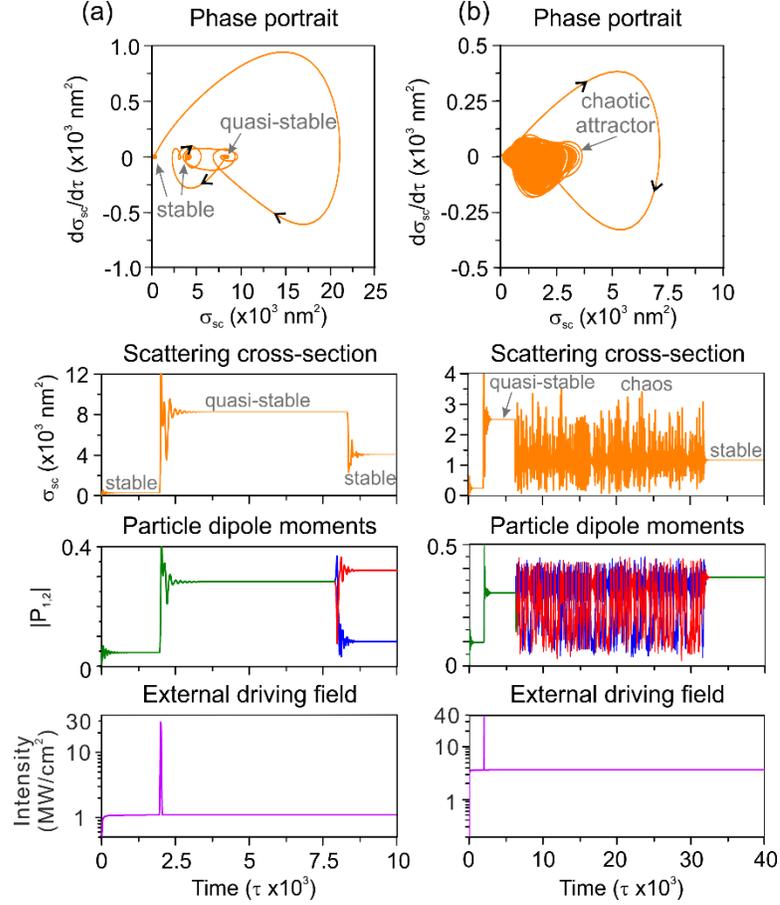

**Fig. 3.** Examples of nonlinear dynamics in the nanodimer associated with the functionality of tristable multivibrator. Bottom panels show the external driving light intensity vs time. The resulted dynamical responses for the slow varying amplitudes of the nanoparticle dipole moments and the nanodimer scattering cross-sections are presented in the second and the third lines from the bottom. The top figures show the corresponding phase portraits for the scattering cross-sections. In the case (a) the system at first comes to the stable symmetric steady state indicated in Fig. 2(c) by the lower green star. Then the signal pulse switches the system to the quasi-stable equilibrium point shown in Fig. 2(c) by the empty star. After a while, modulation instability induced the spontaneous transition to the stable asymmetric steady state (spontaneous symmetry breaking). In (b) the triggering from the quasi-stable to the stable equilibrium point is reached via transitional chaotic dynamics. The dipole moments of the nanoparticles are marked with red and blue when they are different and green when they are equal. For (a) $E_0 = 3.61 \times 10^{-3}$ (1.05 MW cm$^{-2}$), $E_{peak} = 18.61 \times 10^{-3}$ (27.9 MW cm$^{-2}$), $\tau_0 = 2000$ and for (b) $E_0 = 6.76 \times 10^{-3}$ (3.7 MW cm$^{-2}$), $E_{peak} = 21.76 \times 10^{-3}$ (38.15 MW cm$^{-2}$), $\tau_0 = 2000$. For both cases $\Omega = -0.1$ and $\theta_0 = \pi/6$.

Figure 3 (a) demonstrates switching from the symmetric to the asymmetric equilibria via a quasi-stable steady state. This transition is accompanied by the dramatic change in the scattering cross-section, transforming it from almost dark to bright. In this example the duration of keeping the quasi-stable steady state is 26.1 ps which is ~ 290 times longer than the signal pulse (~ 0.09 ps) initiated the first transition. Hence, this functionality can be used to convert short sharp pulses into much longer ones (as in a monostable multivibrator) as well as storing and processing the bits of information (as in a bistable multivibrator).

In the second case, the quasi-stable equilibrium co-exists with 2 stable steady states and a chaotic attractor (Figs. 2(c) and 2(f)). As a result, the transition from the quasi-stable to the eventual stable steady state is performed via a chaotic dynamical process (Fig. 3(b)). Thus, the regime of tristable multivibrator offers an interesting possibility for turning a short signal pulse into comparatively long chaotic pulsations of the dimer scattering cross-section with a fixed duration defined by the system properties.

Figure 4 shows transitions to the regime of an astable multivibrator wherein the system produces periodic self-oscillations. This behavior can be driven by both hard and soft excitation, i.e., with the help of a signal pulse (Fig. 4(a)) and as a result of spontaneous switching from a quasi-stable steady state (Fig. 4(b)), respectively. Additionally, the self-oscillations can be in symmetric and asymmetric modes. Their modulation frequency is much lower than the resonance frequency of nanoparticles and can be tuned from ~ $0.01\omega_0$ to ~ $0.1\omega_0$. Specifically, for the self-oscillations in Figs. 4(a) and 4(b) the modulation frequencies are $0.0065\omega_0$ (0.21 THz) and $0.0142\omega_0$ (0.457 THz), respectively. In this regime the total scattering cross-section is also modulated with quite a high ratio between maximal and minimum values about 3 and 50 for realizations in Figs. 4(a) and 4(b), respectively.

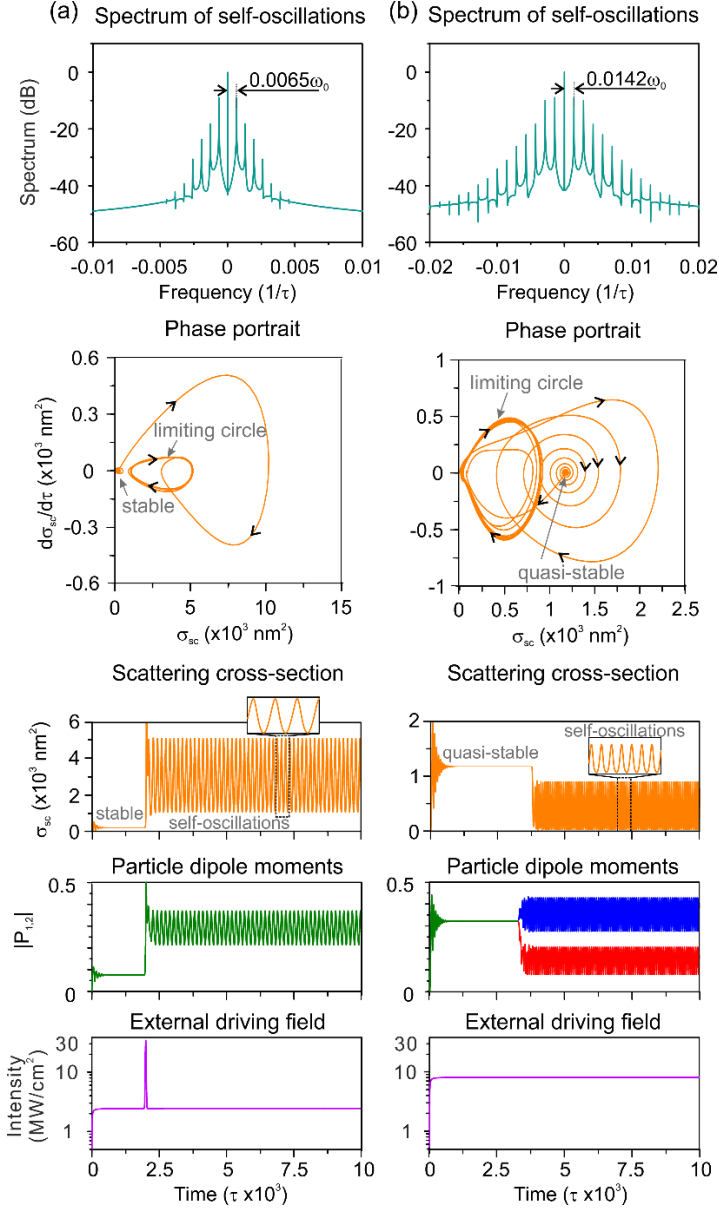

**Fig. 4.** Examples of nanodimer nonlinear dynamics associated with the functionality of an astable multivibrator. In (a) the self-oscillations are induced by the signal pulse (hard excitation) and the symmetry in the system is kept. In (b) the self-oscillations appear spontaneously (soft excitation) as a result of transition from the quasi-stable equilibria and the system symmetry is broken. The spectra are retrieved for the time intervals with self-oscillations. The dipole moments of the nanoparticles are marked with red and blue when they are different and green when they are equal. Insets show the shape of the total scattering cross-section within the time interval $\tau_{int} = 500$. For (a) $\vartheta_0 = \pi/6$, $E_0 = 5.48 \times 10^{-3}$ (2.43 MW cm$^{-2}$), $E_{peak} = 20.48 \times 10^{-3}$ (33.93 MW cm$^{-2}$), $\tau_0 = 2000$ and for (b) $\vartheta_0 = 0$, $E_0 = 10.95 \times 10^{-3}$ (9.72 MW cm$^{-2}$). For both cases $\Omega$=-0.1.

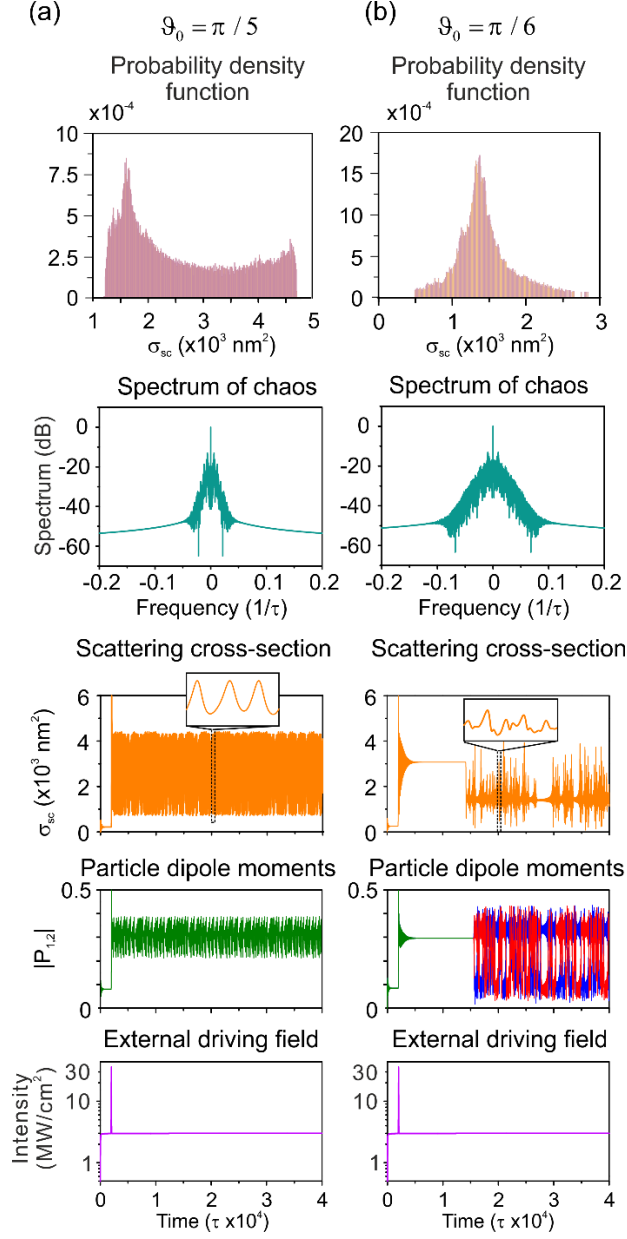

**Fig. 5.** Examples of chaos generation. In (a) the chaotic regime is induced by the signal pulse (hard excitation), and the symmetry in the system is kept. In (b) the system, at first, is driven to take a quasi-stable equilibrium point. Then the chaotic regime is undergoing spontaneously (soft excitation) along with the symmetry breaking. Insets show the shape of the total scattering cross-section within the time interval $\tau_{int} = 500$. The spectra and the probability density functions are calculated for the chaotic regimes at the very long temporal interval $\tau_{int} = 100000$. The dipole moments of the nanoparticles are marked with red and blue when they are different and green when they are equal. For (a) $\vartheta_0 = \pi/5$ and for (b) $\vartheta_0 = \pi/6$. In both cases $E_0 = 6.08 \times 10^{-3}$ (3 MW cm$^{-2}$), $E_{peak} = 21.08 \times 10^{-3}$ (36.06 MW cm$^{-2}$), and $\tau_0 = 2000$.

Next, we consider chaotic dynamical behavior of the nanoantenna, shown in Fig. 5. As in previous cases, the chaotic regime can be accompanied by both preserving (Fig. 5(a)) and breaking (Fig. 5(b)) the system symmetry. The triggering between these two modes can be induced by only small variations in the wave incidence angle (Figs. 5 (a) and (b) obtained for the same set of parameters except for the incidence angle). In contrast to self-oscillations, the Fourier spectra of chaos are continuous and quite broad.

Furthermore, we calculate the probability density functions for the temporal realizations of the scattering cross-sections. To this end, we discretize them with the time step $\tau_{st} = 1$ within the time intervals where chaotic dynamics exist (from $\tau_{min} = 2050$ to $\tau_{max} = 40000$ for Fig. 5(a) and from $\tau_{min} = 14200$ to $\tau_{max} = 40000$ for Fig. 5(b)) and take 500 equidistant sampling points for $\sigma_{sc}$, spanning from its minimal to the maximal value. Then we count the number of values of $\sigma_{sc}$ approximately corresponding to each sampling point and normalize the resulted distribution to one. We characterize the probability density functions by evaluating the mean, the standard deviation, the skewness, and the excess kurtosis as follows: 2640 nm$^2$, 1068 nm$^2$, 0.48, -1.18 for Fig. 5(a) and 1473 nm$^2$, 1694 nm$^2$, 1.56, 3.62 for Fig. 5(b). The comparatively large means and standard deviations with respect to the means offer that these chaotic pulsations are well-pronounced and can be measured in experiment via monitoring the scattered intensity. The positive skewness indicates that the probability density functions have longer right tales than left ones. The excess kurtosis refers to the size of the tails on a distribution with respect to the normal (Gaussian) distribution. Specifically, the negative excess kurtosis for the symmetric chaotic process (Fig. 5 (a)) shows that this process is platykurtic, producing fewer and less extreme outliers than does the normal distribution. In contrast, the asymmetric chaotic signal (Fig. 5 (b)) is leptokurtic characterizing by a positive excess kurtosis, i.e., the tails of the probability distribution function approach zero more slowly than for a Gaussian process, and therefore it produces more outliers than the normal distribution.

Eventually, we stress that the finite skewness and excess kurtosis evidence the non-Markovian (or nonlocal) nature for these stochastic processes. In microelectronics, such complicated dynamical features have been put forward to be used in information processing operations such as confusion and diffusion on the encrypted data and random key generation. However, due to physical limitation on the processing speed, the bitrate of microelectronics devices does not typically exceed ~10 Mb s$^{-1}$. For the nonlinear dimer nanoantenna the available bitrate can be estimated, in accord to the Nyquist–Shannon sampling theorem, from the spectrum bandwidth as ~ $0.01\omega_0$ that corresponds to ~ 2 Tb s$^{-1}$ for a graphene-based nanoantenna operating at the wavelength ~ 9.3 um. This value can be increased further for nonlinear nanoantennas operating in the visible range.

It is instructive to note that observation of the considered functionalities requires quite strong illumination powers, up to ~40 MW cm$^{-2}$ that may cause thermal damage of the nanoantenna associated with changing its optical properties. However, the recent study has shown that coating plasmonic nanoantennas with a $Al_2O_3$ thin layer protects them from thermal damage during the 10 min illumination at the pulse repetition rate 44 MHz and the pulse intensity 0.6 GW cm$^{-2}$ by efficient management of the thermal flow [33]. Additionally, recently there have been published several experimental works showing ultrafast temporal modulation in the scattering of nonlinear optical nanoantennas [34,35] where differential reflectivity measurements allowed to minimize undesired impact of background. Thus, all predicted effects seem readily detectable in experiment.

## 4. Conclusion and Outlook

To summarize, we presented the general analysis of the dynamical behavior for a nonlinear nanoantenna made of a pair of resonant nanoparticles. We have discovered that this simple system can operate in the regimes of tristable and astable multivibrators as well as chaos generator. In contrast to similar devices based microelectronic elements, nonlinear nanoantennas offer much higher operational speeds, that can be used for ultrafast all-optical information processing.

Our findings can be extended to a variety of nonlinear nanophotonic systems, holding great promise for practical use. Specifically, the presented dynamical model can describe nonlinear dynamical behavior in plasmonic [29,36], magnetooptical [37] and high-index [34] nanoparticles, graphene flakes [38] and other nonlinear nanoantennas. The recently developed techniques allowing integration of such elements in silicon nanophotonic circuitry [39] and fiber optics [40] pave the way to numerous opportunities for implementation of nonlinear dynamical functionalities on well-established technological bases.

**Acknowledgement**

This work was supported by ERC StG 'In Motion', PAZY Foundation (Grant No. 01021248), the National Natural Science Foundation of China (Grant No. 11774252), the Qing Lan project, "333" project (Grant No. BRA2015353), PAPD of Jiangsu Higher Education Institutions and China Scholarship Council (CSC).